\documentclass[conference]{IEEEtran}\usepackage{url}
\IEEEoverridecommandlockouts
\usepackage{cite}
\usepackage{amsmath,amssymb,amsfonts}
\usepackage{algorithmic}
\usepackage{graphicx}
\usepackage{textcomp}
\usepackage{hyperref}
\usepackage{booktabs}
\usepackage{multirow}

\usepackage{xcolor}

\usepackage{xspace}
\newcommand{\paragraphX}[1]{\vskip 4pt \noindent \textit{#1} \xspace}

\begin{document}

\title{Evaluating Post-Quantum Cryptographic Algorithms on Resource-Constrained Devices
}


\author{\IEEEauthorblockN{Jesus Lopez$^\dagger$}
\IEEEauthorblockA{
Department of Computer Science\\
University of Texas at El Paso\\
\texttt{jlopez126@miners.utep.edu}}
\and
\IEEEauthorblockN{Viviana Cadena$^\dagger$}
\IEEEauthorblockA{
Department of Computer Science\\
University of Texas at El Paso\\
\texttt{vcadena1@miners.utep.edu}}
\and
\IEEEauthorblockN{Mohammad Saidur Rahman}
\IEEEauthorblockA{
Department of Computer Science\\
University of Texas at El Paso\\
\texttt{msrahman3@utep.edu}}
\thanks{$^\dagger$Equal contribution.}
}

\maketitle

\thispagestyle{plain}
\pagestyle{plain}

\begin{abstract}
The rapid advancement of quantum computing poses a critical threat to classical cryptographic algorithms such as RSA and ECC, particularly in Internet of Things (IoT) devices, where secure communication is essential but often constrained by limited computational resources. This paper investigates the feasibility of deploying post-quantum cryptography (PQC) algorithms on resource-constrained devices. In particular, we implement three PQC algorithms --- BIKE, CRYSTALS-Kyber, and HQC --- on a lightweight IoT platform built with Raspberry Pi devices. Leveraging the Open Quantum Safe (\texttt{liboqs}) library in conjunction with \texttt{mbedTLS}, we develop quantum-secure key exchange protocols, and evaluate their performance in terms of computational overhead, memory usage, and energy consumption for quantum secure communication. Experimental results demonstrate that the integration of PQC algorithms on constrained hardware is practical, reinforcing the urgent need for quantum-resilient cryptographic frameworks in next-generation IoT devices. The implementation of this paper is available at \url{https://iqsec-lab.github.io/PQC-IoT/}.

\end{abstract}

\begin{IEEEkeywords}
Post-Quantum Cryptography; Constrained Devices; IoT Security; Quantum-Secure Communication
\end{IEEEkeywords}

\section{Introduction}

The widespread integration of Internet of Things (IoT) devices into modern infrastructure, from home automation systems to connected vehicles, have introduced new levels of connectivity, convenience, and efficiency. These devices, typically powered by low-cost microcontrollers, operate under stringent resource constraints, including limited processing power, memory, and energy capacity~\cite{6740844}. While this lightweight design supports scalability, it often comes at the cost of robust security~\cite{Morgner_2018, 8796409}. As a result, many consumer-grade IoT devices lack strong cryptographic protections, leaving personal data and network integrity vulnerable to attack.

Securing IoT systems has long been a priority, but the rapid advancement of quantum computing elevates the urgency of this challenge~\cite{thabit2023cryptography}. Classical public-key cryptographic systems such as RSA and elliptic-curve cryptography (ECC), which secure much of today’s digital communication including IoT applications, are vulnerable to quantum attacks~\cite{subramani2025review}. In particular, Shor’s algorithm can efficiently solve the integer factorization and discrete logarithm problems that underpin RSA and ECC, rendering them insecure in the presence of large-scale quantum computers~\cite{shor1999polynomial, bernstein2017post}. This looming threat necessitates the adoption of cryptographic algorithms that remain secure even against quantum adversaries.

Post-quantum cryptography (PQC) addresses this need by developing cryptographic schemes based on mathematical complexity and are resistant to quantum-powered attacks~\cite{bernstein2009introduction, bernstein2017post}. In recognition of the urgent need for standardization, the U.S. National Institute of Standards and Technology (NIST) launched a multi-year effort to identify quantum-resistant cryptographic algorithms~\cite{mosca2018cybersecurity, nist2024final}. In July 2022, NIST announced the first set of finalists: CRYSTALS-Kyber for key encapsulation, and CRYSTALS-Dilithium, Falcon, and SPHINCS+ for digital signatures~\cite{nist2024final}.


In practice, secure communication in embedded systems often relies on Transport Layer Security (TLS) or TLS-inspired protocols, where key encapsulation mechanisms (KEM's) play a central role in negotiating session keys~\cite{rfc8446}. Improving the classical key exchange components in TLS with PQC-based KEM's is one of the most immediate and critical steps in transitioning toward quantum-resistant infrastructure~\cite{alkim2016post}. However, most studies on PQC performance have focused on general-purpose processors, leaving a gap in understanding their behavior in real-world embedded environments~\cite{nist2024final}.

In this paper, we conduct a systematic evaluation of three NIST-designated PQC key encapsulation mechanisms (KEM) --- Bit Flipping Key Encapsulation (BIKE)~\cite{bike}, Hamming Quasi-Cyclic (HQC)~\cite{hqc}, and CRYSTALS-Kyber~\cite{kyber2021spec} on Raspberry Pi-based platforms representative of embedded IoT devices. We measure and compare their performance across four critical dimensions -- execution time, power consumption, memory usage, and device temperature. The goal is to assess the feasibility and efficiency of PQC algorithms for secure communication in resource-constrained environments, and to provide practical insights for future PQC integration in embedded systems.

In summary, the contributions of this paper are as follows:

\begin{itemize}
    \item We present a standardized, reproducible benchmarking methodology for evaluating PQC algorithms using a TLS-inspired client-server model deployed on constrained embedded platforms.

    \item We implement and evaluate three NIST-designated PQC KEM's -- BIKE, HQC, and CRYSTALS-Kyber, at multiple security levels, measuring execution time, power consumption, memory usage, and thermal behavior on Raspberry Pi-based systems.


    \item Our findings show that CRYSTALS-Kyber provides the most favorable trade-off across performance metrics, demonstrating high suitability for secure deployment in constrained IoT environments.

    \item We identify key trade-offs in BIKE and HQC, showing that while BIKE achieves the lowest memory usage, it incurs significant latency and power costs at higher security levels; HQC offers moderate runtime efficiency but suffers from high memory demand and thermal overhead.
\end{itemize}

\section{Related Work}


Research on securing Internet of Things (IoT) devices with post-quantum cryptography (PQC) algorithms has received significant attention from industry and academia due to the growing risks posed by quantum computing~\cite{gharavi2024post, liu2018securing, nist_pqc_2024}. Prior work suggests that public-key schemes such as RSA, DSA, Diffie–Hellman, and ECC can be broken by Shor’s algorithm by efficiently solving integer factorization and discrete logarithm problems~\cite{monz2016realization, shor1999polynomial, shor1994algorithms}. Furthermore, Grover’s algorithm reduces the security of symmetric-key encryption (e.g., AES) and hash-based message authentication codes (e.g., GMAC, Poly1305) by providing a quadratic speedup~\cite{bernstein2017post, sarah2024practical, grover1996fast}. However, symmetric key systems can still be secured by increasing the key sizes, but public-key systems require new designs.


Prior work also emphasized the quantum threats to Internet-of-Things (IoT) devices focusing on threat models, resource usage, models and security analyses for IoT applications~\cite{fernandez2019post, ebrahimi2019post, tasopoulos2023energy}. Liu et al.~\cite{liu2024postquantumcryptographyinternetthings} identified key challenges in integrating post-quantum cryptography (PQC) into IoT, notably limited memory, energy constraints, and the computational overhead of large key sizes, particularly for lattice- and code-based schemes. They conducted a comprehensive survey of over 30 studies on PQC performance, optimization, and GPU acceleration for resource-constrained devices, highlighting Kyber and Dilithium as promising candidates due to their balance between speed and resource consumption. Their work also established benchmark metrics—including speed, code size, transmission size, memory usage, energy, and power—and emphasized the need for a unified evaluation methodology aligned with NIST standardization efforts to ensure consistent benchmarking across constrained IoT platforms.

Recent benchmarking of NIST-selected post-quantum cryptographic (PQC) algorithms on Raspberry Pi 4 devices \cite{fitzgibbon2024benchmarking} evaluated Round 4 winners across lattice-, hash-, code-, and isogeny-based categories, excluding SIKE. The study assessed Kyber, Dilithium, Falcon, SPHINCS+, BIKE, Classic McEliece, and HQC across key generation, encapsulation, signing, verification, and TLS 1.3 handshake performance, showing that Kyber-768 and Dilithium3 achieve superior key exchange and signing efficiency, respectively, while Falcon yields the smallest handshake size. Despite these gains, the need for standardized optimization practices was emphasized \cite{fitzgibbon2024benchmarking}. Further optimization efforts have targeted code-based schemes; for instance, an optimized HQC implementation on an ARM Cortex-M4 platform using ChibiOS RTOS \cite{aissaoui:hal-04699351} achieved 96\% and 95\% reductions in key generation and encapsulation times, respectively, while lowering the memory footprint compared to PQClean. However, further improvements and side-channel resistance evaluations remain necessary \cite{aissaoui:hal-04699351}. In parallel, high-level synthesis (HLS) techniques have been explored for hardware acceleration of NIST Round 3 candidates \cite{Soni2021}, yielding up to 43× and 18× latency reductions for AES and NTT primitives, respectively. Integrating these optimizations into CRYSTALS-Dilithium and CRYSTALS-Kyber resulted in approximately 40\% area savings and 10\% latency improvements, demonstrating that Dilithium2 architectures outperform Dilithium-medium and that HLS provides an efficient alternative to hand-coded RTL for PQC design.

\section{Preliminaries}
To ensure a fair comparison among algorithms, we evaluate only the Key Encapsulation Mechanism (KEM) component. Symmetric encryption of plaintext is performed using the Advanced Encryption Standard in Galois/Counter Mode (AES-GCM), and SHA-256 is employed for key derivation.

\subsection{AES-GCM}

AES-GCM is a widely adopted authenticated encryption algorithm that provides both confidentiality and integrity. It integrates the AES block cipher in Counter (CTR) mode with Galois field multiplication (GHASH) to generate an authentication tag. GHASH compresses the Additional Authenticated Data (AAD) and ciphertext into a single block, ensuring tamper detection~\cite{McGrew2005Galois, dworkin2007recommendation}.

AES-GCM is highly parallelizable, FIPS-approved, and deployed in protocols such as TLS 1.3, IPsec, and SSH~\cite{Kampanakis2024Practical, bhargavan2013implementing}. It performs two operations: authenticated encryption and authenticated decryption.

During authenticated encryption, the algorithm requires:
\begin{itemize}
    \item Secret key: Symmetric encryption key.
    \item Initialization vector (IV): Typically a unique 96-bit nonce.
    \item Plaintext: Data to encrypt.
    \item Optional AAD: Authenticated but unencrypted metadata.
\end{itemize}

The outputs are:
\begin{itemize}
    \item Ciphertext: The encrypted message.
    \item Authentication tag: A fixed-length tag, typically 128 bits.
\end{itemize}

During authenticated decryption, the ciphertext, IV, authentication tag, and optional AAD are used to verify authenticity. Decryption proceeds only if the recalculated tag matches the received tag; otherwise, an error is returned~\cite{McGrew2005Galois}.

Proper IV management is critical: IV reuse under the same key breaks security guarantees~\cite{gueron2013aes}. In this work, IVs are randomized for every encryption operation to mitigate this risk.

\subsection{SHA-256}
SHA-256, standardized in FIPS-180-4, is a cryptographic hash function that produces a 256-bit output from input data of arbitrary length~\cite{fips180-4}. It operates by dividing the input into 512-bit blocks and applying a compression function based on logical operations and modular addition. Designed to offer collision and preimage resistance, SHA-256 also ensures that small changes in input lead to significantly different outputs. These properties make it a reliable choice for deriving symmetric keys from shared secrets in key encapsulation mechanisms (KEMs). Due to its balance of computational efficiency and security, SHA-256 remains well-suited for use in constrained environments such as IoT systems.


\subsection{Naming Conventions}
The National Institute of Standards and Technology (NIST) recommends the following naming conventions for clarity regarding the algorithms discussed in this article: BIKE~\cite{bike}, HQC~\cite{hqc}, and CRYSTALS-Kyber~\cite{kyber2021spec}. Specifically, CRYSTALS-Kyber may also be referred to as Federal Information Processing Standards (FIPS) Publication 203~\cite{NIST.FIPS.203}, as it has been officially standardized. BIKE can only be referenced by its algorithm name, since it is a finalist in the NIST PQC competition but is not selected for standardization. Finally, HQC may be temporarily referred to by its algorithm name, as it has been selected for standardization, but its corresponding FIPS publication has not yet been released at the time of this work~\cite{nist2025hqc}.

\section{Methodology}
This study evaluates the performance of three post-quantum key encapsulation mechanisms (KEMs) -- BIKE~\cite{bike}, CRYSTALS-Kyber~\cite{bos2018crystals}, and HQC~\cite{hqc}, within a simulated client-server communication framework. The goal is to assess the feasibility of deploying these algorithms on constrained embedded systems by measuring their power consumption, execution time, memory usage, and communication throughput.

\begin{figure}[!t]
    \centering
    \includegraphics[width=\linewidth]{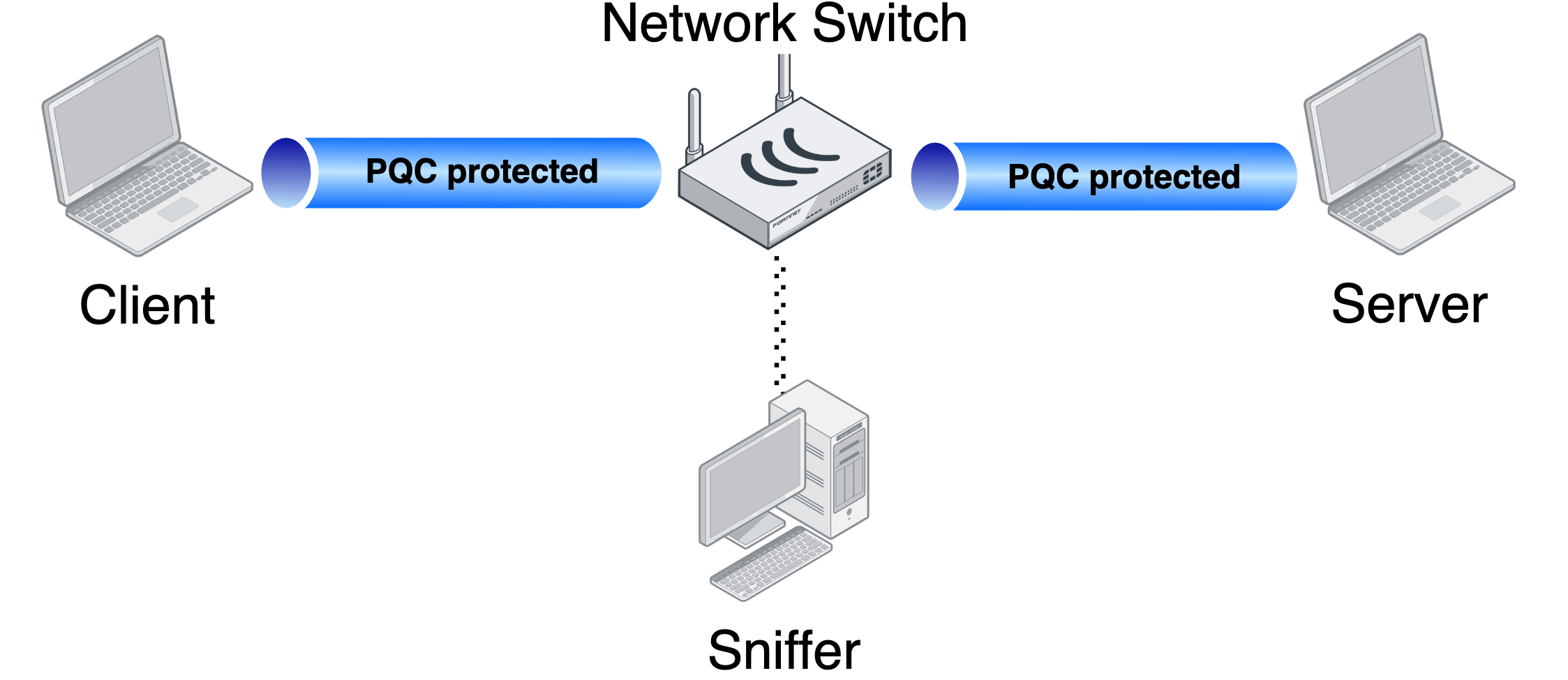}
    \vspace{-0.3cm}
    \caption{Experimental setup for PQC-protected communication between a client and a server. A sniffer is connected through a network switch to monitor traffic flow and passively verify correct message routing. }
    \label{fig:experimental-setup}
    \vskip -0.3cm
\end{figure}

\subsection{Experimental Setup}

The experimental setup, shown in \autoref{fig:experimental-setup}, consisted of a Raspberry Pi 3 Model B+ as the client and a Raspberry Pi 5 as the server. Both devices ran Raspberry Pi OS Lite, a Debian-based system selected for its minimal overhead and lightweight design. An offline network switch connected the two devices, providing an isolated environment free from external interference to ensure consistent and reproducible measurements.


All software is developed in C, using mbedTLS (version 3.6) as the primary cryptographic library~\cite{mbedtls}, extended with liboqs from the Open Quantum Safe (OQS) project (version 0.12)~\cite{liboqs}. Each KEM is integrated independently into the communication protocol to allow for a fair evaluation.



Testing includes three parameter sets for each KEM to assess performance at varying security levels. BIKE is evaluated with BIKE-L1, BIKE-L3, and BIKE-L5; HQC with HQC-128, HQC-192, and HQC-256; and CRYSTALS-Kyber with Kyber512, Kyber768, and Kyber1024. Each one of the levels align with the NIST PQC standardization goals~\cite{nist-pqc-cfp-2016} of providing 128-bit, 192-bit and 256-bit key. These levels mirror classical cryptographic strength AES-128, AES-192/SHA-384, AES-256/SHA-512. This selection enables a consistent comparison of the algorithm families across different cryptographic strengths.

A passive network sniffer, connected to the switch, monitors client-server communications. Packet exchanges are captured using Wireshark over a wired interface~\cite{wireshark}, ensuring accurate observation without introducing overhead or altering the communication behavior.



\begin{figure}[!t]
    \centering
    \includegraphics[width=\linewidth]{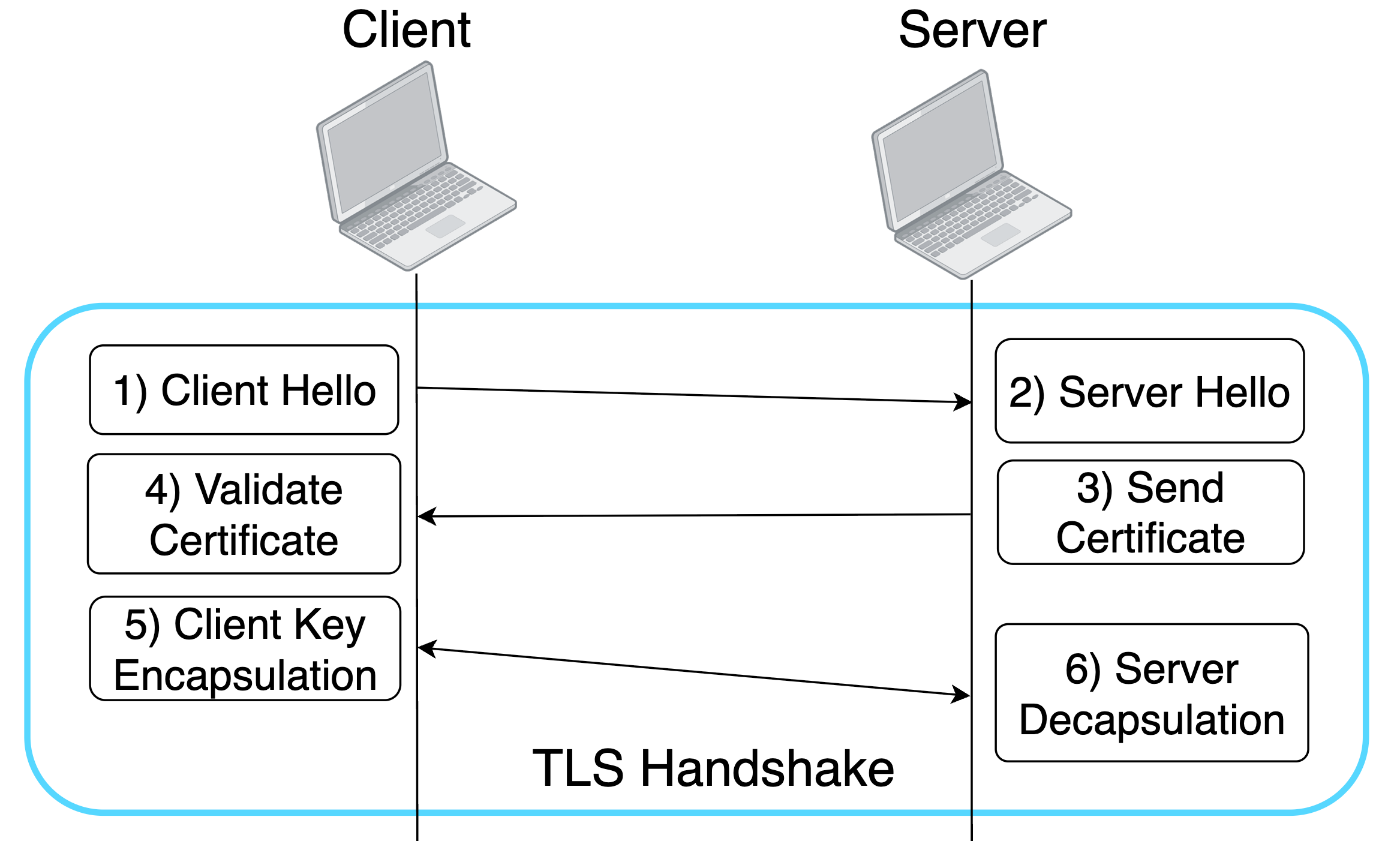}
    \vspace{-0.3cm}
    \caption{Simplified TLS handshake protocol, illustrating the exchange between the client and the server. }
    \label{fig:communication-protocol}
    \vskip -0.3cm
\end{figure}


\subsection{Implementation Procedure}

A custom lightweight protocol is implemented for benchmarking, consisting of the following steps:
\begin{itemize}
    \item Handshake: The client initiates a secure session by establishing a TCP connection with the server, shown in \autoref{fig:communication-protocol}.
    \item Key Exchange:
    \begin{itemize}
        \item The server generates a key pair using the selected post-quantum key encapsulation mechanism (PQC-KEM) and sends the public key to the client.
        \item The client performs key encapsulation and returns the resulting ciphertext and shared secret to the server.
    \end{itemize}
    \item Session Key Derivation: Both the client and server derive a common session key from the KEM output.
    \item Secure Message Exchange: To validate successful key agreement, the client encrypts a message using AES-GCM and transmits it to the server.
\end{itemize}


\subsection{Performance Evaluation Procedure}
\label{sec:performanceevaluate}



To evaluate the performance of each algorithm, the client and server applications are executed concurrently in isolated sessions. Four text files of varying sizes (208 bytes, 731 bytes, 1235 bytes, and 2328 bytes) are used to simulate different communication loads and observe the impact of message size on system performance. The use of text files were chosen for their simplicity and transparency, allowing us to easily control the data size, format, and structure. This approach is straightforward for manual testing, reproducibility and consistency. Each session runs for a fixed duration of five minutes, followed by a five-minute rest period to allow the devices to return to baseline thermal and performance conditions.

Performance data is collected continuously during each session. The following metrics are recorded for analysis:
\begin{itemize}
    \item Total execution time (seconds)
    \item Power consumption (watts)
    \item Memory usage (kilobytes)
    \item Device temperature (degrees Celsius)
\end{itemize}

In the case of power consumption, we are measuring the server and the client under low-load conditions, which means that the system is idle mainly or under minimal stress. In contrast, High-load conditions represent when the system is under stress; we used this to measure the peak power consumption of a typical use in a real-world application.


\section{Evaluation}

\subsection{Results}

\begin{figure}[!t]
    \centering
    \includegraphics[width=\linewidth]{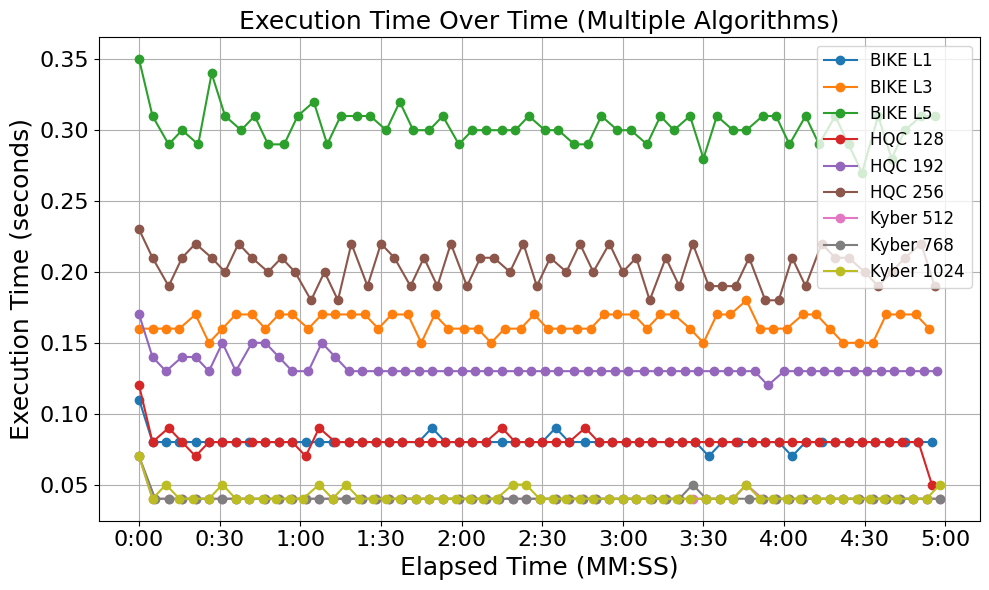}
    
    \vspace{-0.3cm}
    \caption{Execution time comparison of post-quantum key encapsulation mechanisms (KEMs) over a five-minute interval. The three levels of BIKE, HQC, and Kyber algorithms were evaluated across multiple security levels.}
    \label{fig:execution_time}
    \vspace{-0.3cm}
\end{figure}

\begin{table}[!t]
\centering
\caption{Execution Time (in seconds) for Post-Quantum Cryptography Algorithms.}
\label{tab:meanTimeandStandardDeviation}
\begin{tabular}{l|c}
\toprule
\textbf{PQC Algorithm} & \textbf{Execution Time} \\
\midrule
BIKE L1 & 0.081 $\pm$ 0.005  \\
BIKE L3 & 0.164 $\pm$ 0.007  \\
BIKE L5 & 0.302 $\pm$ 0.013  \\
\hline
HQC 128 & 0.081 $\pm$ 0.007  \\
HQC 192 & 0.133 $\pm$ 0.008  \\
HQC 256 & 0.203 $\pm$ 0.013  \\
\hline
Kyber 512 & 0.041 $\pm$ 0.004  \\
Kyber 768 & 0.041 $\pm$ 0.004  \\
Kyber 1024 & 0.042 $\pm$ 0.005  \\
\bottomrule
\end{tabular}
\vspace{-0.3cm}
\end{table}

\begin{table}[t]
\centering
\caption{Power consumption (Watts) of PQC Algorithms.}
\label{tab:PQCpowerConsumption}
\begin{tabular}{l|cc|cc}
\hline
\toprule
& \multicolumn{2}{c|}{\textbf{Low}} & \multicolumn{2}{c}{\textbf{High}} \\ \cline{2-5}
\textbf{PQC Algorithm} & \textbf{Client} & \textbf{Server} & \textbf{Client} & \textbf{Server} \\
\midrule
BIKE L1      &3.2 &3.5 &3.5 & 4.7 \\
BIKE L3      &3.2 &3.5 &3.5 &4.8 \\
BIKE L5      &3.2 &3.5 &3.5 &5.2 \\
\hline
HQC 128      &3.2 &3.4 &3.6 &4.0 \\
HQC 192      &3.2 &3.5 &3.6 &4.1 \\
HQC 256      &3.2 &3.5 &3.7 &4.7 \\
\hline
Kyber 512      &3.2 &3.4 &3.5 &3.9 \\
Kyber 768      &3.2 &3.5 &3.5 &3.9 \\
Kyber 1024     &3.2 &3.5 &3.5 &3.9 \\
\bottomrule
\end{tabular}
\vspace{-0.3cm}
\end{table}

\begin{table*}[!t]
\centering
\caption{Memory Usage (in kilobytes) of PQC Algorithms in Four Network Scenarios.}

\label{tab:Meanstdnetsim1}
\begin{tabular}{c|c|c|c|c}
\toprule
\textbf{Algorithm} & \textbf{networkSim1}& \textbf{networkSim2} & \textbf{networkSim3} & \textbf{networkSim4} \\
\midrule
BIKE L1 & 5632.00 $\pm$ 0.00  & 5632.00 $\pm$ 0.00 & 5632.00 $\pm$ 0.00 &  5629.79 $\pm$ 16.81\\
BIKE L3 &  5627.59 $\pm$ 33.61 & 5629.79 $\pm$ 16.81&  5620.97 $\pm$ 43.45 &  5629.79$\pm$ 16.81\\
BIKE L5 & 5851.43 $\pm$ 93.63  & 5888.00 $\pm$ 0.00 &  5858.29 $\pm$ 80.92 &  5858.29 $\pm$ 84.52\\
\hline
HQC 128 & 5757.83 $\pm$ 16.66 & 5760.00 $\pm$ 0.00 & 5749.15 $\pm$ 35.95 & 5753.49 $\pm$ 28.36\\
HQC 192 & 5998.34 $\pm$ 55.97 & 5987.31 $\pm$ 68.00& 6000.55 $\pm$ 48.42 & 6000.55 $\pm$ 59.11 \\
HQC 256 & 5874.53 $\pm$ 46.43 & 5881.26 $\pm$ 28.84 & 5879.02 $\pm$ 32.99 & 5883.51 $\pm$ 23.76 \\
\hline
Kyber 512 & 5888.00 $\pm$ 0.00 & 5885.83 $\pm$ 16.66 & 5888.00 $\pm$ 0.00  & 5760.00  $\pm$ 0.00\\
Kyber 768 & 5760.00 $\pm$ 0.00 & 5757.83 $\pm$ 16.66 & 5757.83 $\pm$ 16.66 &5885.83 $\pm$ 16.66 \\
Kyber 1024 & 5888.00 $\pm$ 0.00 & 5888.00 $\pm$ 0.00  & 5888.00 $\pm$ 0.00 & 5883.66 $\pm$ 33.33 \\
\bottomrule
\end{tabular}
\vspace{-0.3cm}
\end{table*}

\begin{table*}[t]
\centering
\caption{Mean and Standard Deviation of Temperatures on Server and Client Devices (in $^\circ$C)}
\label{tab:tempsCombined}
\begin{tabular}{c|cccc|cccc}
\toprule
\multirow{2}{*}{\textbf{Algorithm}} & \multicolumn{4}{c|}{\textbf{Server Temperature ($^\circ$C)}} & \multicolumn{4}{c}{\textbf{Client Temperature ($^\circ$C)}} \\
\cline{2-9}
 & networkSim1 & networkSim2 & networkSim3 & networkSim4 & networkSim1 & networkSim2 & networkSim3 & networkSim4 \\
\midrule
BIKE L1 & 48.37$\pm$0.88 & 50.55$\pm$0.95 & 50.83$\pm$0.93 & 51.86$\pm$1.02 & 48.88 $\pm$0.59 & 49.77$\pm$0.42 & 50.45$\pm$0.42 & 52.64$\pm$0.48 \\
BIKE L3 & 51.04$\pm$1.32 & 51.33$\pm$1.09 & 51.11$\pm$1.27 & 52.26$\pm$1.09 & 52.69$\pm$0.36 & 52.60$\pm$0.48 & 52.39$\pm$0.46 & 52.26$\pm$0.42 \\
BIKE L5 & 51.53$\pm$1.29 & 51.75$\pm$1.30 & 52.04$\pm$1.26 & 51.84$\pm$1.09 & 53.41$\pm$0.47 & 53.40$\pm$0.49 & 53.38$\pm$0.51 & 53.51$\pm$0.38 \\
\hline
HQC 128 & 48.23$\pm$0.88 & 49.16$\pm$1.03 & 49.49$\pm$0.90 & 51.70$\pm$0.74 & 53.29$\pm$0.40 & 53.09$\pm$0.55 & 53.48$\pm$0.49 & 53.66$\pm$0.42 \\
HQC 192 & 48.43$\pm$0.95 & 49.05$\pm$0.90 & 50.81$\pm$1.21 & 51.31$\pm$1.08 & 53.86$\pm$0.46 & 54.20$\pm$0.52 & 53.92$\pm$0.48 & 54.08$\pm$0.45 \\
HQC 256 & 50.22$\pm$0.85 & 50.07$\pm$1.01 & 50.53$\pm$1.08 & 51.11$\pm$0.90 & 53.90$\pm$0.48 & 52.21$\pm$0.64 & 52.41$\pm$0.60 & 52.44$\pm$0.46 \\
\hline
Kyber 512 & 47.96$\pm$0.87 & 47.54$\pm$1.29 & 49.03$\pm$1.20 & 50.01$\pm$1.16 & 52.15$\pm$0.44 & 52.15$\pm$0.53 & 51.98$\pm$0.48 & 52.35$\pm$0.49 \\
Kyber 768 & 47.27$\pm$0.70 & 47.62$\pm$0.97 & 48.92$\pm$1.06 & 49.66$\pm$1.16 & 52.09$\pm$0.47 & 52.14$\pm$0.54 & 52.20$\pm$0.48 & 51.97$\pm$0.46 \\
Kyber 1024 & 47.85$\pm$1.06 & 47.62$\pm$1.19 & 49.27$\pm$1.07 & 49.81$\pm$1.08 & 52.21$\pm$0.46 & 52.24$\pm$0.40 & 51.95$\pm$0.47 & 52.06$\pm$0.48 \\
\bottomrule
\end{tabular}
\vspace{-0.3cm}
\end{table*}

This section presents the performance evaluation of three post-quantum key encapsulation mechanisms (KEMs) -- BIKE, HQC, and Kyber -- across different security levels. 


\subsubsection{Execution Time}
The execution time of each algorithm is evaluated over continuous five-minute sessions, with metrics recorded at each iteration as described in~\autoref{sec:performanceevaluate}.

\autoref{fig:execution_time} shows the execution time behavior of all evaluated algorithms. Among the families tested, the Kyber variants consistently exhibit the lowest and most stable execution times. Kyber-512, Kyber-768, and Kyber-1024 report mean execution times of 0.041 $\pm$ 0.004 seconds, 0.041 $\pm$ 0.004 seconds, and 0.042 $\pm$ 0.005 seconds, respectively, as summarized in \autoref{tab:meanTimeandStandardDeviation}. The small observed fluctuations, with standard deviations between $\pm$0.004 and $\pm$0.005 seconds, indicate that Kyber maintains a highly consistent runtime, with minimal sensitivity to external conditions or input variations.

In contrast, the BIKE family shows significantly higher execution times. BIKE-L1 and BIKE-L3 report mean execution times of 0.081 $\pm$ 0.005 seconds and 0.164 $\pm$ 0.007 seconds, respectively, while BIKE-L5 achieves the highest latency at 0.302 $\pm$ 0.013 seconds. 

The HQC family demonstrates intermediate performance. HQC-128 matches BIKE-L1 with an execution time of 0.081 $\pm$ 0.007 seconds, while HQC-192 and HQC-256 report 0.133 $\pm$ 0.008 seconds and 0.203 $\pm$ 0.013 seconds, respectively. The higher standard deviations observed for BIKE-L5 and HQC-256 ($\pm$0.013 seconds) reflect greater variability compared to Kyber variants.





\subsubsection{Power Consumption}


Power consumption is measured separately under low-load and high-load conditions for both the client and the server,as shown in \autoref{tab:PQCpowerConsumption}. Low-load corresponds to idle and High-load simulates peak usage scenarios representative of real-world applications.

Under low-load conditions, all algorithms demonstrate similar behavior, consuming approximately 3.2 Watts on the client side. Server-side consumption varies slightly between 3.4 to 3.5 Watts across all algorithms. Under high-load conditions, differences become more pronounced. BIKE-L5 records the highest server-side power consumption at 5.2 Watts, followed by BIKE-L3 at 4.8 Watts and BIKE-L1 at 4.7 Watts. On the client side, all BIKE variants maintain 3.5 Watts.

The HQC family shows moderate high-load power consumption. HQC-128 and HQC-192 consume 4.0 Watts and 4.1 Watts on the server side, respectively, while HQC-256 rises slightly to 4.7 Watts. Client-side power consumption for HQC variants ranges from 3.6 to 3.7 Watts.

The Kyber family maintains the lowest power consumption under high-load conditions, with Kyber-512, Kyber-768, and Kyber-1024 each consuming approximately 3.5 Watts on the client and 3.9 Watts on the server.




\subsubsection{Memory Usage}
Memory usage for each algorithm is summarized in \autoref{tab:Meanstdnetsim1}, measured across four different communication scenarios (networkSim1 to networkSim4) corresponding to increasing file sizes. The four input files contain increasingly larger text-based inputs, with sizes of 208 bytes, 731 bytes, 1235 bytes, and 2328 bytes, respectively. This setup allows us to evaluate how our measurements changes with the change of input size, simulating varying communication loads and scenarios.


In networkSim1 (208 bytes), BIKE-L1 records a mean memory usage of 5632.00 $\pm$ 0.00 KB, while BIKE-L3 uses 5627.59 $\pm$ 33.61 KB. These are the lowest among all tested algorithms. In contrast, HQC-192 and Kyber-512 reach the highest memory usage, reporting 5998.34 $\pm$ 55.97 KB and 5888.00 $\pm$ 0.00 KB, respectively. Kyber-1024 similarly reports 5888.00 $\pm$ 0.00 KB.

As the file size increases in networkSim2 (731 bytes), networkSim3 (1235 bytes), and networkSim4 (2328 bytes), overall memory usage remains relatively stable for most algorithms. HQC-192 consistently shows the highest memory usage, ranging between 5987.31 $\pm$ 68.00 KB and 6000.55 $\pm$ 59.11 KB across the four simulations. BIKE-L1 and BIKE-L3 continue to demonstrate the lowest memory usage, maintaining values between 5620.97 $\pm$ 43.45 KB and 5632.00 $\pm$ 0.00 KB.

Variability in memory usage is reflected in the standard deviation values. BIKE-L5 displays the greatest fluctuation, with deviations reaching up to $\pm$ 93.63 KB in networkSim1. In comparison, Kyber-512 and Kyber-1024 show no measurable fluctuation in several simulations, indicating highly stable memory consumption.

Overall, BIKE-L1, BIKE-L3, and Kyber-512 demonstrate both low and stable memory usage, while HQC-192 shows higher and more variable memory demand across all file sizes.




\subsubsection{Device Temperature}
Temperature measurements for both client and server devices are reported in \autoref{tab:tempsCombined}.

For server temperatures, the Kyber family consistently records the lowest operating temperatures across all scenarios. Kyber-768 maintains the lowest average temperatures, ranging from 47.27$^\circ$C to 49.66$^\circ$C across the four network simulations. Kyber-512 and Kyber-1024 report similar thermal profiles, with temperatures between 47.54$^\circ$C and 50.01$^\circ$C.

The BIKE and HQC families demonstrate slightly higher operating temperatures. BIKE-L1, BIKE-L3, and BIKE-L5 show average server temperatures exceeding 50$^\circ$C under load, with BIKE-L5 reaching up to 52.04$^\circ$C in networkSim3. Similarly, HQC-256 records temperatures ranging from 50.07$^\circ$C to 51.11$^\circ$C across different simulations.

Client-side temperatures follow a similar trend. Kyber variants again maintain lower thermal profiles compared to BIKE and HQC. Some missing data points are observed in the client-side measurements for BIKE-L3 and BIKE-L5, likely due to inconsistencies during data collection. Nonetheless, the overall trend aligns with server-side observations, highlighting Kyber’s thermal efficiency relative to BIKE and HQC.

%
%

\begin{figure}[!t]
    \centering
    \includegraphics[width=\linewidth]{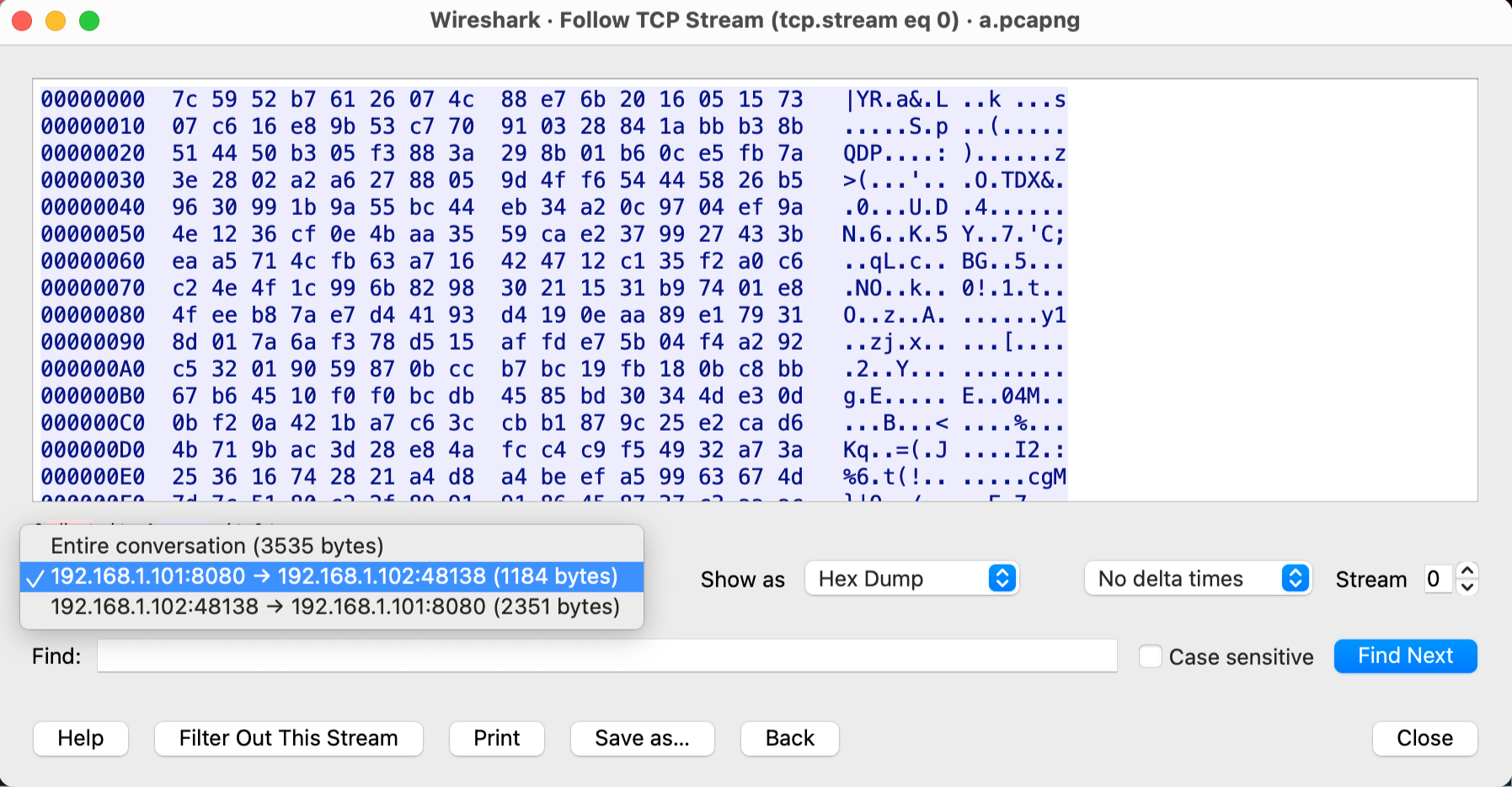}
    \vspace{-0.3cm}
    \caption{Wireshark output showing the captured TCP stream. The highlighted packet is 1,184 bytes in size, which corresponds to the expected ciphertext length produced by the Kyber-768 PQC algorithm.}
    \label{fig:wireshark}
    \vspace{-0.3cm}
\end{figure}

\subsubsection{Packet Sniffer}
A sniffer was placed between Alice and Bob, as illustrated in \autoref{fig:experimental-setup}, to monitor the communication channel of client and server. ARP spoofing is employed to redirect and capture network packets. Since PQC protocols are not yet integrated into mainstream protocols such as TLS, Wireshark is currently unable to natively recognize them. Given that our implementation uses PQC within a custom protocol, protocol-specific identification is not feasible using standard packet analyzers. Therefore, analysis is performed by directly inspecting the contents of the transmitted packets.

\autoref{fig:wireshark} depicts, as an example, the use of the Kyber-768 algorithm to demonstrate a specific packet in the network trace. The size of the transmitted message is 1,184 bytes, which corresponds to the expected ciphertext length generated by the Kyber-768 key encapsulation mechanism.






\subsection{Analysis}
The evaluation results highlight significant performance differences across the three post-quantum key encapsulation mechanisms (KEMs): BIKE, HQC, and Kyber.

Across all metrics, Kyber consistently outperforms the other algorithms. Kyber variants achieve the lowest execution times, with Kyber-512 and Kyber-768 operating at 0.041 $\pm$ 0.004 seconds and Kyber-1024 at 0.042 $\pm$ 0.005 seconds. They also demonstrate minimal runtime variability. Correspondingly, Kyber maintains the lowest high-load power consumption, approximately 3.5 Watts on the client side and 3.9 Watts on the server side, and the lowest operating temperatures, ranging between 47.27$^\circ$C and 50.01$^\circ$C. Memory usage for Kyber variants is slightly higher than BIKE but remains highly stable, with almost no fluctuation in several scenarios. These results indicate that Kyber offers a balanced profile of computational efficiency, low energy consumption, and thermal stability, making it highly suitable for constrained embedded systems.

BIKE shows mixed performance. While BIKE-L1 and BIKE-L3 maintain relatively low memory footprints (around 5630 KB) and low idle power consumption, their execution times are significantly higher compared to Kyber. BIKE-L5, in particular, exhibits the highest latency (0.302 $\pm$ 0.013 seconds) and the highest server-side power consumption under load (5.2 Watts). BIKE variants also show higher thermal profiles, with BIKE-L5 reaching up to 52.04$^\circ$C, and increased variability in memory usage, particularly for BIKE-L5. These trends suggest that while BIKE may be competitive at lower security levels, its scalability to higher security parameters introduces notable performance penalties.

HQC demonstrates intermediate behavior across most metrics. HQC-128 offers execution times comparable to BIKE-L1 (0.081 $\pm$ 0.007 seconds), but HQC-192 and HQC-256 show increases in memory consumption (up to 6000.55 $\pm$ 59.11 KB) and power consumption (server-side power reaching 4.7 Watts for HQC-256). HQC variants also experience elevated operating temperatures, similar to BIKE. The increased memory demand and variability, especially at higher security levels, may pose challenges for resource-constrained platforms.

Overall, the Kyber family provides the most balanced and efficient performance across all tested metrics. BIKE and HQC offer viable alternatives depending on application-specific constraints but exhibit trade-offs in execution time, energy consumption, and memory stability, particularly at higher security levels.

\section{Discussion}

While post-quantum cryptography offers strong security guarantees, its deployment on resource-constrained devices remains non-trivial. Our results reveal that CRYSTALS-Kyber, a lattice-based scheme built on the Module Learning with Errors (M-LWE) problem~\cite{kyber2021spec}, consistently outperforms code-based alternatives BIKE and HQC in execution time, energy efficiency, and thermal stability. Kyber’s computational advantages and implementation maturity~\cite{ni2023hpka} make it particularly well-suited for embedded environments.


Despite optimization efforts, HQC and BIKE exhibit higher resource demands than Kyber. Hardware-accelerated designs for HQC~\cite{aissaoui:hal-04699351} and constant-time implementations of BIKE for embedded platforms~\cite{chen2021bike} improve performance but do not fully address their elevated energy and time costs. Our results confirm these trends: both schemes show increased energy consumption and memory usage, particularly at higher security levels, limiting their suitability for constrained environments.

All three algorithms maintain acceptable power usage on the client side, but BIKE-L1 and BIKE-L3 are preferable for systems with strict memory constraints, consistently demonstrating the lowest memory consumption across all test scenarios. To ensure fair comparison, all experiments are conducted under consistent conditions using identical operating systems (Raspberry Pi OS Lite v6.6), codebases differing only by the integrated KEM, and an isolated network to eliminate external interference. A Raspberry Pi 5 functions as the server, and a Raspberry Pi 3 as the client, modeling practical resource asymmetry. No significant differences are observed in handling varying file sizes or transmission durations. Packet loss and delays remain negligible due to the controlled testbed and small payload sizes.

In summary, while all evaluated algorithms are functional on resource-constrained platforms, CRYSTALS-Kyber offers the most efficient trade-off across execution time, power consumption, memory usage, and thermal behavior. BIKE and HQC remain viable alternatives under specific application constraints but present notable scalability challenges as security levels increase. These insights are critical for guiding PQC algorithm selection in future embedded and IoT deployments.

\section{Conclusion}

In this study, we practically evaluate the performance of three post-quantum key encapsulation mechanisms -- BIKE, HQC, and CRYSTALS-Kyber, on resource-constrained devices. Experimental results show that CRYSTALS-Kyber consistently achieves superior efficiency in execution time, power consumption, memory usage, and thermal behavior, making it the most suitable candidate for embedded and IoT platforms. While BIKE and HQC remain viable under specific conditions, their higher resource demands, particularly at increased security levels, present scalability challenges. These findings highlight the critical need to balance cryptographic strength with system-level constraints when selecting PQC algorithms for future deployments at scale.

\paragraphX{\textbf{Acknowledgments.}} Jesus Lopez is funded by the Department of Computer Science at UTEP. Viviana Cadena is funded by the Cyber Halo Innovation Research Program (CHIRP) at Pacific Northwest National Laboratory (PNNL).

\bibliographystyle{IEEEtran}
\bibliography{ref}

\end{document}